\documentclass[twocolumn,showpacs,preprintnumbers,amsmath,amssymb]{revtex4}
\usepackage{graphicx}% Include figure files
\usepackage{dcolumn}% Align table columns on decimal point
\usepackage{amsmath}
\usepackage{amsfonts}
\usepackage{bm}% bold math

\def\beq{\begin{equation}}
\def\eeq{\end{equation}}

\def\phi{{\varphi}}
\def\equal{\buildrel {\rm def} \over {=} }
\def\={ \equal } 
\newcommand{\Mag}{\mathcal{M}}
\newcommand{\vect}[1]{\mathbf{#1}}
\DeclareMathOperator{\erfc}{erfc}
\begin{document}

\preprint{APS/123-Cond Mat}

\title{Transition to chaos in magnetized, weakly coupled plasmas}

\author{Andrea Carati}
 \email{andrea.carati@unimi.it}
\affiliation{Department of Mathematics, Universit\`a degli Studi di   Milano\\
         Via Saldini 50, 20133 Milano, Italy}
\author{Francesco Benfenati}
 \email{francesco.benfenati@studenti.unimi.it}
\affiliation{Corso di Laurea in Fisica, Universit\`a degli Studi di   Milano\\
         Via Celoria 12, 20133 Milano, Italy}
\author{Alberto Maiocchi}
 \email{alberto.maiocchi@unimi.it}
\author{Luigi Galgani}
 \email{luigi.galgani@unimi.it}
\affiliation{
    Department of Mathematics, Universit\`a degli Studi di   Milano\\ 
    Via Saldini 50, 20133 Milano, Italy}
\author{Matteo Zuin}
\email{matteo.zuin@igi.cnr.it}
\affiliation{Consorzio RFX, Associazione EURATOM-ENEA sulla Fusione, 
Padova, Italy }

\date{\today}

\begin{abstract}
We report  the results of numerical simulations  for a
model of a one component plasma (a system  of $N$ point electrons with
mutual Coulomb  interactions) in a uniform stationary  magnetic
field. 
We take $N$ up to
512, with periodic boundary conditions, and macroscopic  parameters corresponding
to the weak coupling regime,  with  a coupling parameter 
$\Gamma=1/64$. 
We find that a transition from order to
chaos takes place
when  the density is increased  or the field decreased so that  the
ratio $\omega_p/\omega_c$ between plasma and cyclotron
frequencies becomes of order $1$ (or equivalently  the ratio $r_L/\lambda_D$ between
Larmor radius and Debye length  becomes of order  $1$). 
The result is in agreement with the theoretical prediction
obtained  in   \cite{plasmichaos}, on the basis of  an    
 old estimate
of Iglesias, Lebowitz and MacGowan\cite{lebo} for the intensity of
the  electric field acting on one electron and due to all the other
ones. A comparison can be made
with the threshold  obtained from  
kinetic theory arguments, which  corresponds to the condition 
$\nu_{ee}/\omega_c=1$, where $\nu_{ee}$ is  the electron collision
frequency. The latter threshold has a completely different dependence
on the physical parameters and, for  $\Gamma=1/64$, gives a critical
value of $\omega_p$ about  $80$ times larger.
\end{abstract}

\maketitle

The microscopic foundations of plasma physics are usually formulated
in terms of kinetic theory, in which a key role is played by the concept of
``collision frequency''  or by the related one of ``mean free path''. 
It is usually
stated that for the so--called  weakly coupled plasmas, such as gaseous--discharge
plasmas, fusion plasmas, or a plasma in the solar corona,   the Coulomb
coupling is so small that ``\emph{their
  thermodynamic properties are analogous  to those of an ideal
  gas}'' (see \cite{ichimaru}, page 10), i.e.,  the electrons
behave essentially as if they were free. 
%Correspondingly, for a weakly
%coupled plasma immersed in an external uniform
%stationary magnetic field the transverse  motions
%are apparently expected to  be just slight perturbations of the unperturbed pure
%Larmor  gyrations.

A different  approach was taken in  paper \cite{plasmichaos}, in
which the microscopic Newton equations  themselves were tackled
directly,  making use of the tools of  ergodic theory and of a
quite recent extension  of Hamiltonian perturbation theory to the
thermodynamic limit (see  \cite{car1,maicar,carmai}). This allows one
to obtain theoretical results for the microscopic model itself, 
with no  need of passing through the
approximation of the Boltzmann equation. In particular, for  
an infinite plasma immersed in a uniform stationary
 magnetic field $\vect B$ the electron motions were
estimated to be ordered 
if the ratio  $\omega_p/\omega_c$ between electron plasma and
cyclotron frequencies  is below unity, chaotic in the 
opposite case. This means that the Coulomb interactions among the
electrons are strong enough to produce  a chaoticity  threshold  when
$\omega_p\simeq\omega_c$ (the definitions of these and of some other familiar
quantities will be recalled in a moment).
 Here ``ordered'' has to be understood in the sense of
ergodic theory, i.e., that there exists at least one dynamical
variable the 
time--autocorrelation of which does not decay to zero, or decays in an
extremely slow way. For completely chaotic motions, instead, 
the time--correlations of smooth dynamical variables are known to
quickly decay  to zero. 

Notice that the estimate for the chaoticity  threshold determined in
\cite{plasmichaos} can
be eventually  expressed in an extremely intuitive way, namely, as the condition
that the  typical value of the perturbing force on any electron
(the sum of the Coulomb forces due to all the other ones) just equals the
typical value of the Lorentz force. So, denoting by $E_j$ the modulus
of the electric field acting on the $j$--th electron
and by $v_j$ the modulus of the electron's velocity, the condition for
the chaoticity threshold takes the
simple form $E_j\simeq Bv_j/c$ (in Gauss units).  In turn, the
electric field acting on one electron and due to all the other
ones,  looked at as a random variable,
obviously has vanishing mean,  so that  its typical value  is
estimated by its standard deviation. The value of the latter, at
density $n_e$ and temperature $T$, was estimated long ago by Iglesias,
Lebowitz and MacGowan \cite{lebo} to be given by $\sqrt{4\pi\, n_e k_B T}$,
where $k_B$ is
the Boltzmann constant. This leads  for the threshold
to the condition 
$\omega_p/\omega_c\simeq 1$, which in particular is independent of the coupling
parameter $\Gamma$.

We  used here the definitions 
$\omega_c\= eB/mc$,  $\omega_p\=\sqrt{e^2 n_e/m}$, $\Gamma\= e^2/ ak_BT$,
where  $m$ and $e$ are the electron mass and charge,  $c$ the speed
of light, and   $a\=n_e^{-1/3}$ the mean interparticle distance. 
A relevant related quantity is the Debye length
$\lambda_D\=\sqrt{k_BT/n_e e^2}$.

The  theoretical estimate  for the  chaoticity
threshold  at $\omega_p\simeq \omega_c$ found in 
\cite{plasmichaos} was quite unexpected because  kinetic
arguments apparently suggest that, in the weakly coupled regime $\Gamma\ll
1$, a transition might occur at $\nu_{ee}\simeq \omega_c$, where
$\nu_{ee}$ is the \emph{electron collision frequency} (see for example
\cite{meiss}). On the other
hand one has  (see for example  \cite{ichimaru}, page 35)
$\nu_{ee}\simeq  \Gamma^{3/2}\,  |\log \Gamma|\, \omega_p$, and this
gives a threshold at $\omega_p\simeq \omega_c/ \left(\Gamma^{3/2}
|\log \Gamma |\right)$, i.e., at $\omega_p\gg\omega_c$ (for $\Gamma\ll 1$).

In this letter we report the results of numerical simulations 
at $\Gamma=1/64$. 
A transition to chaotic motion is seen to occur for
$\omega_p/\omega_c$ in the interval between $0.25$ and $2$,  in
agreement with the prediction given in  \cite{plasmichaos}.

Let us recall that  a one component plasma model is just a system of
$N$  point electrons with mutual Coulomb interactions. We denote
their position vectors by $\vect  x_j$, $j=1,\ldots,N$,  and 
take them to lie in a box of side $L$ (so that the electron density
is given by $n_e=N/L^3$). They are subject to the
Lorentz force $(e/c) \; \vect B \wedge \vect{ \dot
x}_j$ due to a constant homogeneous magnetic
field $\vect B$ (which we take directed along the $z$ axes), and to their mutual
Coulomb forces. The electric  force on the $j$--th electron, which
depends on the positions  $\vect x_1,\ldots,\vect x_N$ of all
electrons, may be simply denoted by $e^2 \vect E(\vect x_j)$, where 
$\vect E$ is the electric field acting on that electron and due unit
charges located in the positions of the other
electrons.  As we are using periodic boundary conditions (so that we are
actually dealing with a system of infinitely many electrons), 
the latter field can be
computed \cite{ewald} by the Ewald summation of the  field due  to an
infinite cubic lattice of charges of the form $\vect x_i + L\vect n$. Here,
$\vect n$ is a vector with integer coordinates, i.e.,
$\vect n\=(n_x \vect e_x+ n_y \vect e_y + n_x\vect e_z)$, with $n_x$, $n_y$ and
$n_z\in\mathbb{Z}$, while $\vect e_x, \cdots$ are the unit vectors along
thr axes.

Rescaling  time  by the electron cyclotron frequency $\omega_c$
and   position vectors by  the mean interparticle
distance $a$, i.e., introducing $\tau \= \omega_c t$
and $\vect y_j\=\vect x_j/a$,  the equations of motion take the form
$$
\ddot{\vect y}_j = \vect e_z\wedge \dot{\vect y}_j +
\left(\frac{\omega_p}{\omega_c}\right)^2 \,  \vect E(\vect y_j) \ 
$$
(the  dots denoting now derivatives with respect to $\tau$),  and so
contain only one (dimensionless) parameter, namely,
${\omega_p}/{\omega_c}$,
while the rescaled density is obviously equal to $1$.
In the simulations, the explicit form of the Ewald resummed field
$\vect E$ acting on the $j$--th particle is given
by
\begin{equation*}
  \begin{split}
    \vect E(\vect y_j) = & \sum_{\vect n}\sum_l \frac {\vect r_{l,\vect
    n}}{|r_{l,\vect n}|^3} \Big[ \erfc(\alpha r_{l,\vect n}) +
      \frac{\alpha r_{l,\vect n}}{\sqrt{\pi}}\exp(-\alpha^2r_{l,\vect
        n}^2) \Big] \\
    &+ \frac {4\pi}{L^3}\sum_{\vect k\ne0}\sum_l \frac{\vect k}{|\vect k|^2}
    \exp(-\frac{\vect k^2}{4\alpha})\sin(\vect k\cdot \vect r_l)\ .
  \end{split}
\end{equation*}
Here  $\vect r_l\= \vect y_j-\vect y_l$, while $\vect r_{l,\vect n}\= \vect
y_j-\vect y_l + L\vect n/a$; the function $\erfc(x)$ is the usual
error function, and $\alpha$ is the Ewald convergence parameter which
we chose as $\alpha \= \pi^{1/2}N^{1/6}L^{-1}$. In the
first sum the term corresponding to the self-force on the $j$--th
particle should be excluded.
\begin{figure*}[ht]
  \begin{center}
    \includegraphics[width=0.45\textwidth]{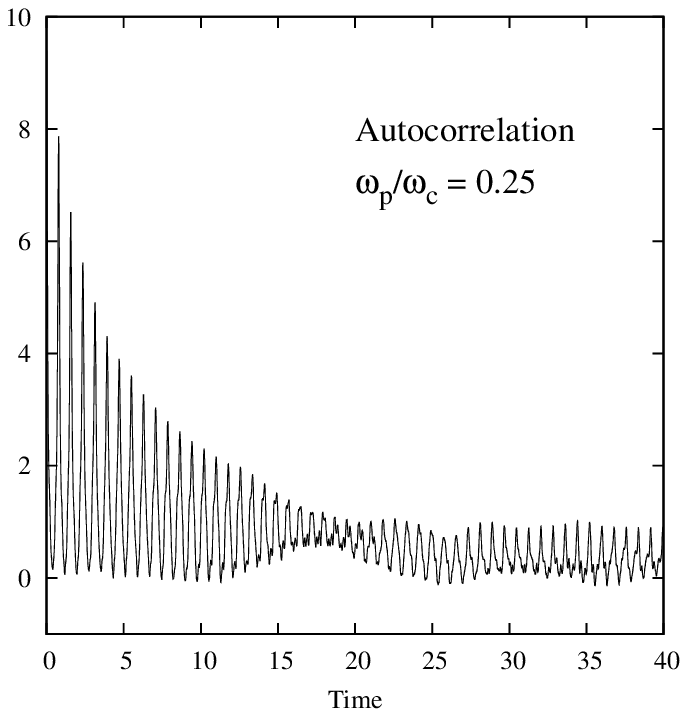}
    \includegraphics[width=0.45\textwidth]{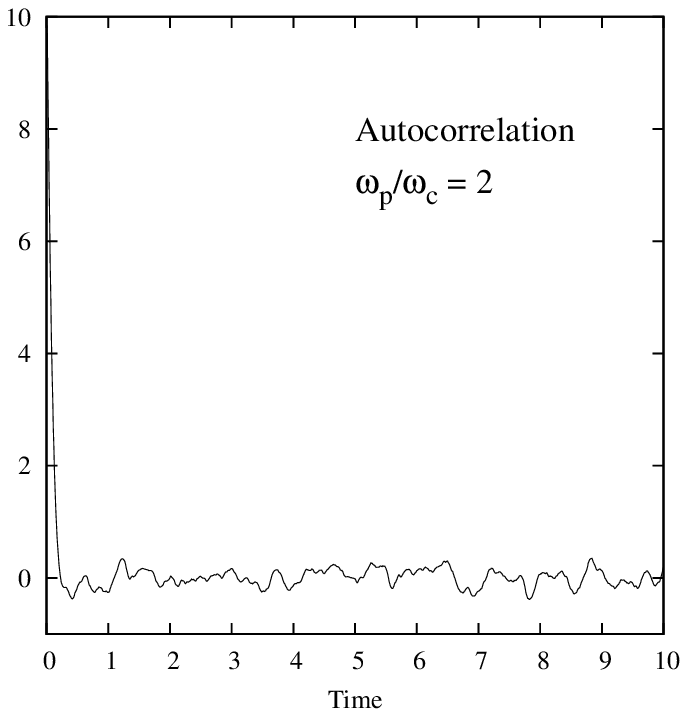} \\
\vskip 0.5 truecm    
    \includegraphics[width=0.45\textwidth]{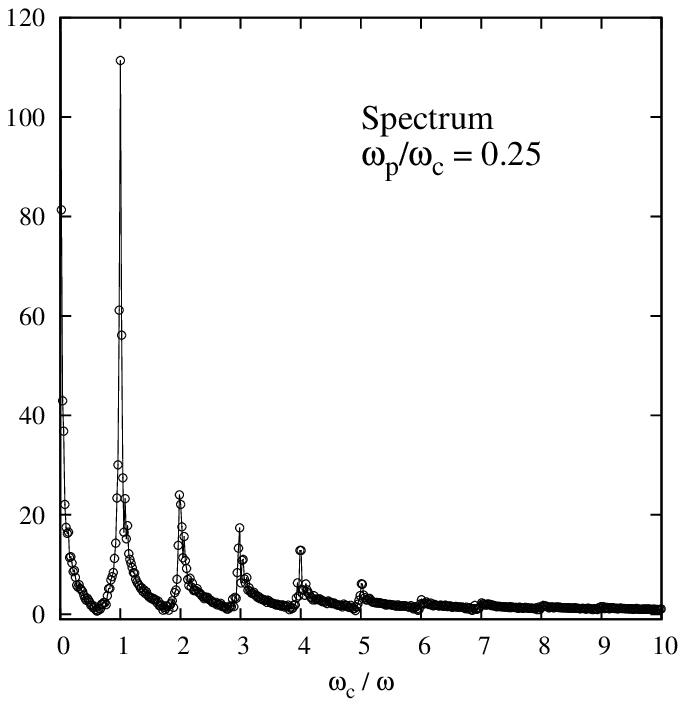}
    \includegraphics[width=0.45\textwidth]{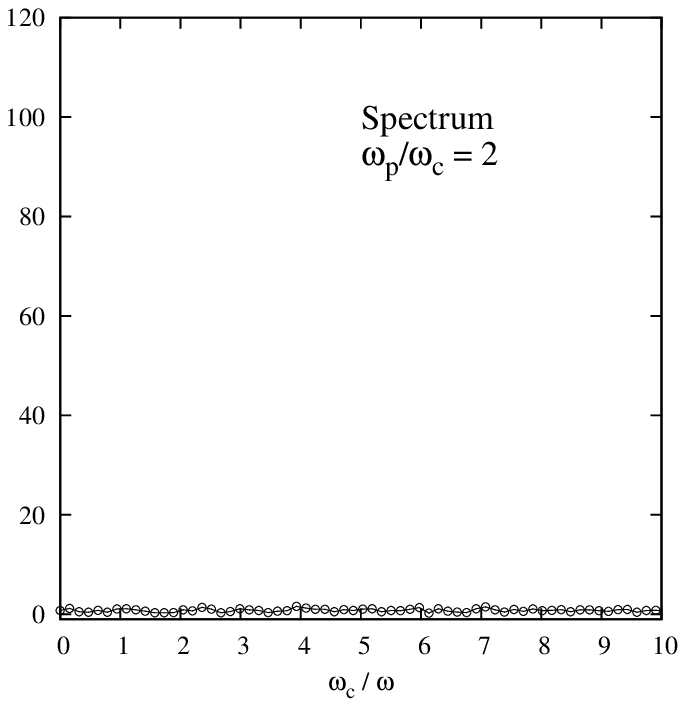}
  \end{center}
  \caption{\label{fig1} Top: Autocorrelation $\mathcal{C}_{\Mag}(t)$  of
    magnetization versus  time for
   $\omega_p/\omega_c=0.25$ (left) and $\omega_p/\omega_c=2$
    (right). The time scale is the same in both figures, and one has
    $\omega_c=1$ at the right, $\omega_c=8$ at the left.
    Notice the fast decay to zero  at
    the right. Bottom: Discrete Fourier transform (absolute value)  of 
    $\mathcal{C}_{\Mag}(t)$ versus $\omega/\omega_c$  
    for $\omega_p/\omega_c=0.25$ (left),  and  $\omega_p/\omega_c=2$
    (right). Peaks (and thus also magnetization)
    have disappeared at the right. Here $\Gamma=1/64$.}
\end{figure*}

These are the equations of motion that were actually  integrated
numerically, using  a symplectic splitting method.
The conservation of energy in every run was better than a part over $10^3$. 
The integration time was chosen proportional to $\omega_c$ in order that
all different cases be integrated for the same
physical time. In any case,   the time was always
some hundreds cyclotron periods.

The initial data were chosen in the following way: 
the electron positions $\vect y_j$ were taken
uniformly distributed in the box of side $N^{1/3}$, while the velocities
were extracted from a Maxwellian with a given temperature $T$. 
This introduces in the model  (in addition to
$\omega_p/\omega_c$) the  further parameter $T$, or equivalently the
dimensionless Coulomb coupling parameter $\Gamma$, to be used in
the Maxwell distribution for the velocities.

For what concerns the number $N$ of electrons in the box,
our computational power allows us to go up to $N=512$. This
induces a lower bound on $\Gamma$, namely, $\Gamma \ge
N^{-2/3}$. Indeed, in order to correctly simulate the 
Coulomb  cumulative force acting on an electron in a plasma, 
the side of the box has to be
at least equal to the Debye length, which, in our rescaled units,
takes the value $\lambda_D=\Gamma^{-1/2}$. We took $\Gamma
=N^{-2/3}$. Computations were performed both for  $N=128$ and $N=512$, which
correspond to $\Gamma=128^{-2/3}\simeq 0.04$ and
$\Gamma=512^{-2/3}=1/64\simeq 0.016$ respectively. 
%Only the
%results for $\Gamma=1/64$ are illustrated  below, because the figures
%corresponding to such two cases are qualitatively similar.

We now come  to the main issue, i.e.,  whether the motions are
ordered or chaotic.
Obviously  what plays the role of the unperturbed system with
completely ordered motions  is the limit case with $\omega_p/\omega_c=0$, for
which the Coulomb interaction disappears and one has pure Larmor gyrations. The
problem  then is to determine whether a threshold for chaotic motions
takes place as the parameter $\omega_p/\omega_c$ is increased and
$\Gamma$ is varied. To this end we 
considered the   magnetization of a box,
$\Mag\= (e/2mc) \sum \vect {\dot x}_j\wedge \vect x_j$, looking at its 
autocorrelation function  (normalized by $Nk_BT$)
$$
 \mathcal{C}_{\Mag}(t)\= \frac {<\Mag(0)\Mag(t)>}{Nk_B T}\ , 
$$
and at its Fourier transform $\hat {\mathcal{C}}_{\Mag}(\omega)$. The
latter  is a physically  very 
relevant quantity  because, according to linear response theory (see
\cite{kubo,klimontovich}, or
Appendix~B of \cite{benfcargal}),  
$i\omega\hat {\mathcal{C}}_{\Mag}(\omega)$  
gives the  susceptibility $\chi(\omega)$ at frequency $\omega$.
In the  formula for the time--autocorrelation
$\mathcal{C}_{\Mag}(t)$, the average $<\cdot >$ is meant  as
a  phase--average with respect  to  Gibbs measure; in our
computations, however,  we estimated it by the time--average along an
orbit (with initial data extracted as previously explained), as often  done in
numerical works. We did not investigate the relations between the two
averages. Moreover, the  Fourier transform $\hat
{\mathcal{C}}_{\Mag}(\omega)$ was estimated by the amplitude of the
discrete Fourier transform of $\mathcal{C}_{\Mag}(t)$, which will be
simply called the spectrum. So we report figures of the
time--autocorrelation $\mathcal{C}_{\Mag}$
versus $t$, and of the corresponding spectrum versus angular frequency
$\omega/\omega_c$.

Having fixed  $\Gamma=1/64$, by increasing $\omega_p/\omega_c$ we found  that
a threshold occurs for $\omega_p/\omega_c$ between $0.25$ and $2$. This is
exhibited in Fig.~1, where  the results are reported for such two values of   
$\omega_p/\omega_c$,  $0.25$ on the left and $2$ on the right. The
autocorrelations are reported in the upper part of the figure, and the
spectra in the lower part.

For $\omega_p/\omega_c=0.25$
the autocorrelation is seen to
display regular oscillations with a decreasing amplitude: we were
unable to follow this relaxation process up to the end. The
oscillations are apparently peaked about the cyclotron frequency and
its low harmonics (as should be, due to the nonlinearities in the equations of
motions). This is clearly exhibited by the 
 spectrum, with its large  peak at $\omega/\omega_c=1$, and the smaller
 ones  about the low harmonics $\omega/\omega_c=2,3,...$. 
Of special relevance is the peak at $\omega=0$, which corresponds to  the
  existence of a nonvanishing static susceptibility, i.e., to the
  existence of diamagnetism.
There also appears
 a continuous component, which accounts for the extremely slow drift towards
equilibrium.  This case  clearly  corresponds to prevalently ordered
motions with a corresponding nonvanishing diamagnetism, 
and should be interpreted as an indication that the
perturbation due to the Coulomb interactions is not yet sufficiently
large to produce  prevalent chaotic motions. 
The passage to chaos,
however,  already occurred at $\omega_p/\omega_c=2$.
Indeed  in this case the autocorrelation is seen to  go to zero in an
extremely short lapse of
time (even shorter than one cyclotron  period $2\pi/\omega_c$), so that
the peaks disappear from the spectrum and  one only  remains with  the continuous
part. 
This means that for $\Gamma=1/64$ the threshold in $\omega_p/\omega_c$ lies
between $0.25$ and $2$.  
For $\Gamma=128^{-2/3}$ the corresponding figures at those same values
of  $\omega_p/\omega_c$ are qualitatively similar to  the above ones,
and  are not reported here.

So, the  numerical results obtained for $\Gamma=128^{-2/3}\simeq
0.04$ and $\Gamma=1/64\simeq 0.016$ 
 are  in rather good
agreement with the theoretical prediction  found in
\cite{plasmichaos}, namely: at $\omega_p/\omega_c =1$ the interactions
become  strong enough as to make the motions chaotic. 
On the other hand this  is  apparently in contrast    with 
 kinetic theory arguments, according to which  
Coulomb interactions should be negligible up to values of $\omega_p$
larger by a factor $20$ and  $80$ respectively.  The discrepancy
would become enormous in physically relevant cases, as
   gaseous--discharge plasmas,  fusion plasmas, or a plasma in the
solar corona, for which $\Gamma$ takes  the  typical  values $10^{-3},10^{-5},
10^{-7}$ respectively. Indeed, in terms of densities, for
fusion plasmas 
the coupling should  be negligible up to densities about thirteen
orders of magnitudes larger than according to the law 
$\omega_p/\omega_c=1$.
 Notice  by the way that, as shown in
 \cite{plasmichaos}, the latter threshold
 appears to fit pretty well, at least as orders of magnitude are
 concerned,  the empirical data for disruptions in  fusion machines.
 
In conclusion,  the present numerical work 
confirms, for weakly coupled plasmas,   the theoretical predictions
for the chaoticity threshold given in \cite{plasmichaos}. The main
point is however that this confirms  the estimate given in
\cite{lebo} for the intensity of  the electric field acting on one
electron and due to all the other ones,  which turns out to be  much
larger than usually assumed.
Now, the fact that the collective  Coulomb effects are relevant  for  
weakly coupled plasmas (with
$\Gamma\ll 1$) is very well known (see for example
\cite{boyd}, page 8). Indeed it is
just for such plasmas  that
the number  $\Gamma^{-3/2}$ of effectively interacting electrons,
(those   contained in a
Debye sphere)  turns out to be  very large. 
What is apparently lacking,
perhaps,  is a general  acquaintance with  how large such an effect 
may actually be, in fact so huge as to possibly explain the disruptions in fusion
machines.  

Such an acquaintance  might perhaps 
 help elucidating  also the  situation met in the problem of  anomalous
transport, where it occurs that \emph{``... measured energy transport
  rates typically exceed those calculated for binary  collisions 
...''} \cite{anomalo2}, or even \emph{``... greatly exceed''} them
\cite{anomalo1}. We hope to come back to this point in the future.
For a study on anomalous diffusion in a  strongly
coupled one component plasma, through numerical computations of the
same type as those  performed here, see  \cite{ottbonitz}.

\textbf{Acknowledgments.} The present paper is dedicated to Francesco
Guerra (La Sapienza University at Rome) on the occasion of his
seventieth birthday.

\end{document}